\documentclass[10pt,
							 twocolumn,
							 superscriptaddress,
							 english,
							 prl,
							 showpacs,
							 floatfix,
							 aps
							]{revtex4-1}
\usepackage[utf8]{inputenc}
\usepackage{amsmath}
\usepackage{amssymb}
\usepackage{graphicx}
\usepackage{xspace}
\usepackage{siunitx}

\makeatletter
\@ifundefined{textcolor}{}
{%
 \definecolor{BLACK}{gray}{0}
 \definecolor{WHITE}{gray}{1}
 \definecolor{RED}{rgb}{1,0,0}
 \definecolor{GREEN}{rgb}{0,1,0}
 \definecolor{BLUE}{rgb}{0,0,1}
 \definecolor{CYAN}{cmyk}{1,0,0,0}
 \definecolor{MAGENTA}{cmyk}{0,1,0,0}
 \definecolor{YELLOW}{cmyk}{0,0,1,0}
}

\usepackage{soul}
\usepackage{braket}
\usepackage[backref=none,
bookmarksnumbered=true,
bookmarks=true,
bookmarksopen=true,
colorlinks=true,
citecolor=blue,
linkcolor=blue,
anchorcolor=green,
urlcolor=blue,unicode=false]{hyperref}
\let\baraccent=\= % rename builtin command \= to \baraccent
\renewcommand{\=}[1]{\stackrel{#1}{=}} % for putting numbers above =

\newcommand{\unitspace}{~}

\newcommand{\Fig}[1]{Fig.\unitspace\ref{fig:#1}}
\newcommand{\Figure}[1]{Figure\unitspace\ref{fig:#1}}

\makeatother

\usepackage{babel}
\begin{document}

\title{Visualizing Intramolecular Distortions as the Origin of Transverse Magnetic Anisotropy}

\author{Daniela Rolf}
\affiliation{Fachbereich Physik, Freie Universit\"at Berlin, 14195 Berlin, Germany}
\author{Christian Lotze}
\affiliation{Fachbereich Physik, Freie Universit\"at Berlin, 14195 Berlin, Germany}
\email{c.lotze@fu-berlin.de}
\author{ Constantin\ Czekelius}
\affiliation{Institut f\"ur Organische Chemie und Makromolekulare Chemie,
 Heinrich-Heine-Universit\"at D\"usseldorf, 40225 D\"usseldorf, Germany}
\author{Benjamin W. Heinrich}
\affiliation{Fachbereich Physik, Freie Universit\"at Berlin, 14195 Berlin, Germany}
\author{Katharina J. Franke}
\affiliation{Fachbereich Physik, Freie Universit\"at Berlin, 14195 Berlin, Germany}

\date{\today}

\begin{abstract}

The magnetic properties of metal-organic complexes are strongly influenced by conformational changes in the ligand. The flexibility of Fe-tetra-pyridyl-porphyrin molecules leads to different adsorption configurations on a Au(111) surface. By combining low-temperature scanning tunneling spectroscopy and atomic force microscopy, we resolve a correlation of the molecular configuration with different spin states and magnitudes of magnetic anisotropy. When the macrocycle exhibits a laterally-undistorted saddle shape, the molecules lie in a S=1 state with axial anisotropy arising from a square-planar ligand field. If the symmetry in the molecular ligand field is reduced by a lateral distortion of the molecule, we find a finite contribution of transverse anisotropy. Some of the distorted molecules lie in a S=2 state, again exhibiting substantial transverse anisotropy.

\end{abstract}

\pacs{%
      %75.70.Rf 	Surface magnetism
			%74.25.Jb, % Superconcuctivity: Electronic structure (photoemission, etc.)
			%74.55.+v % SC: Tunneling phenomena: single particle tunneling and STM
			} %  73.22.-f 	Electronic structure of nanoscale materials and related systems
\maketitle 
%\maketitle must follow title, authors, abstract and \pacs

%%############################ INTRODUCTION ###############################

Transition-metal atoms placed in anisotropic environments such as molecular ligands possess distinct magnetic properties. An anisotropic atomic surrounding leads to the lifting of d-orbital degeneracies, with the spin state of the atom being determined by the interplay between Hund`s coupling and the corresponding d-level splitting. Spin-orbit interactions introduce magneto-crystalline anisotropies, thus favoring particular spin orientations. Engineering the interplay of level splitting and spin-orbit interactions, thus opens the path for designing specific magnetic   properties \cite{rau_reaching_2014,Ben_FeOEP_2015,hiraoka_single_molecule_2017,ormaza_controlled_2017,
chen_abrupt_2018}. 

Metal-organic complexes, such as metal-porphyrins are ideal candidates to manipulate the energies of d-level splitting and magnetic anisotropy, because they provide a large intramolecular flexibility. Indeed, Fe-porphyrins on surfaces have been reported to carry different magnetic ground states  \cite{wang_intramolecularly_2015,
carmen_rubio_verdu_orbital_selective_2017,liu_large_2017}. The studies report different magnetic anisotropy energies and even different total spin states. The origin of the different magnetic properties most probably lies in the different molecular distortions as a response to different molecular arrangements. However, direct evidence of the correlation of intramolecular structure with magnetic states is challenging, as it requires the sensitivity to structural as well as magnetic properties at the level of an individual molecule. 

Here, we use Fe-5,10,15,20-tetra-4-pyridyl-porphyrin (FeTPyP) on a Au(111) surface as a model system to study intramolecular conformations and magnetic properties by a combination of low-temperature atomic force microscopy and tunneling spectroscopy. 
%%%%%%%%%%%%%%%%%%%%%%%%%%%%%%%%%%%%%%%%%%%%%%%%%%%%%%%%%%%%%%%%%%%%%%%%%%%%%%%%%%%%%%%%%%%%%%%%%%%%%%%%%%%%%%%%%%%%%%%%%%%%%%%%%%%%%%%%%%%%%%%%%%%%        Molecule + Topos      %%%%%%%%%%%%%%%%%%%%%%%%%%%%%%%%%%%%%%%%%%%%%%%%%%%%%%%%%%%%%%%%%%%%%%%%%%%%%%%%%%%%%%%%%%%%%%%%%%%%%%%%%%%%%%%%%%%%%%%%%%%%%%%%%%%%%%%%%%%%%%%%%%%%%%%%%%%%%

\begin{figure}[ht]
\includegraphics[width=0.95\columnwidth]{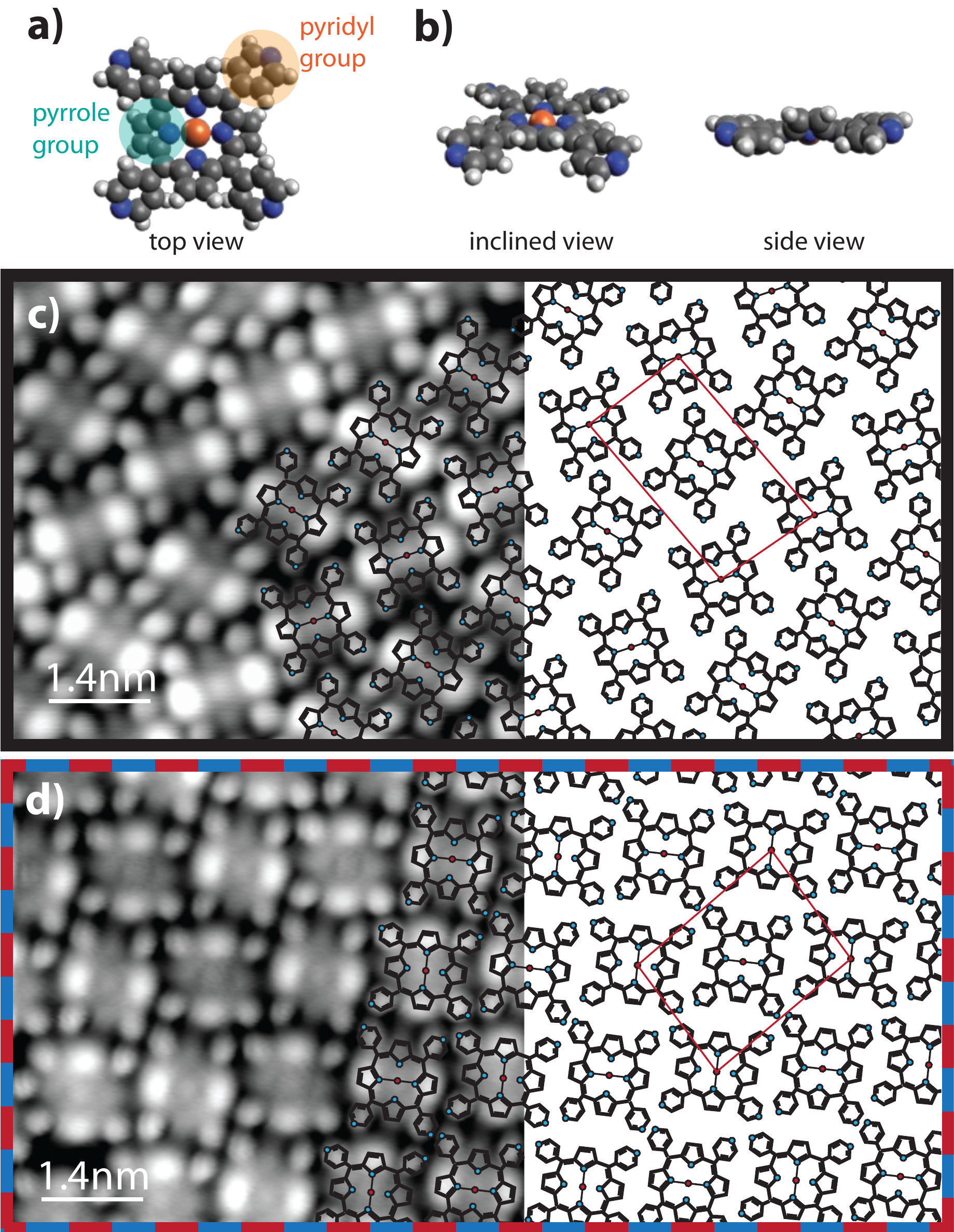}
\caption{a) Structure model of the FeTPyP molecule, indicating the freely rotatable pyridyl groups. b) Inclined and side view of a schematic model of FeTPyP, showing the out-of-plane saddle-shape deformation of the pyrrole groups. c,d) STM topography of two arrangements of FeTPyP on Au(111): c) Densely-packed arrangement of FeTPyP molecules, consisting of  alternating rows of parallel molecules. The red box shows the rhombic unit cell of $\SI{1.4}{\nm} \times \SI{2.7}{nm}$, containing two molecules. d) Staggered arrangement of FeTPyP with an alternating pattern of the molecules. The unit cell of $\SI{2.0}{\nm} \times \SI{2.2}{nm}$ also contains two molecules. Due to the saddle-shape deformation of the molecules, two pyrrole groups of the molecules in both structures appear higher in the STM topography than the other two. The color of the boxes around the images corresponds to the colors of the spectra in Fig.\,\ref{Fig2} and Fig.\,\ref{Fig3}. Topographies recorded at \SI{200}{\pico\A}, \SI{200}{\mV} (c) and \SI{160}{\pico\A}, \SI{230}{\mV} (d), respectively. }
\label{Fig1}
\end{figure}

The FeTPyP molecule consists of a porphyrin core with four rotatable pyridyl groups attached to it (see Fig.\,\ref{Fig1}a). In gas phase, the pyridyl groups are oriented orthogonal to the porphyrin core in order to minimize steric repulsion between the hydrogen atoms of the pyridyl group and the neighboring pyrrole ring.
In response to interactions with the surface, the pyridyl groups tend to adapt a flattened structure upon deposition on a substrate, which - due to  steric hindrance between adjacent hydrogen atoms - induces an out-of-plane distortion of the inner macrocycle. In the resulting saddle-shape geometry, one pair of opposing pyrrole rings bends up and the other pair bends down (see Fig.\,\ref{Fig1}b) \citep{auwarter_controlled_2007,xianwen_chen_conformational_2017}.
Because of this flexibility several molecular arrangements with varieties of intramolecular configurations form on surfaces. This allows us to compare different distortions and the resulting magnetic properties.

The FeTPyP-Cl complex \cite{Fetpyp,Fetpyp2} was evaporated from a Knudsen cell evaporator at \SI{410}{\celsius} with the sample held at room temperature. Upon deposition, the molecules are dechlorinated, with the Fe changing its oxidation state from +3 to +2 \cite{heinrich_change_2013,Ben_FeOEP_2015,gopakumar_transfer_2012}.
Figure \ref{Fig1}c,d show the formation of two self-assembled structures of FeTPyP molecules on the Au(111) substrate.  In both of these structures, the FeTPyP molecules lie in the aforementioned saddle-shape configuration, as can be inferred by the twofold symmetry of the porphyrin macrocycle. Similar shapes have been observed for many metal-tetra-phenyl-porphyrins (M-TPP) and metal-tetra-pyridyl-porphyrins (M-TPyP) on surfaces \citep{auwarter_controlled_2007, albrecht_direct_2016,wang_intramolecularly_2015, auwarter_conformational_2007, xianwen_chen_conformational_2017,zotti_ab-initio_2007, liu_large_2017,carmen_rubio_verdu_orbital_selective_2017}. 

In the first arrangement, referred to in the following as the densely-packed arrangement (Fig.\,\ref{Fig1}c), the molecules arrange in alternating rows. The rhombic unit cell with side lengths of $\SI{1.4}{\nm} \times \SI{2.7}{nm}$ consists of two molecules. The molecular structure as well as the size of the unit cell coincide with the assembly of FeTPyP on Ag(111) \citep{auwarter_controlled_2007}. Figure \ref{Fig1}d shows a second structure of FeTPyP, where the molecules are in a staggered  arrangement and exhibit an alternating orientation of their saddle shape.

\begin{figure}[ht]
\includegraphics[width=0.95\columnwidth]{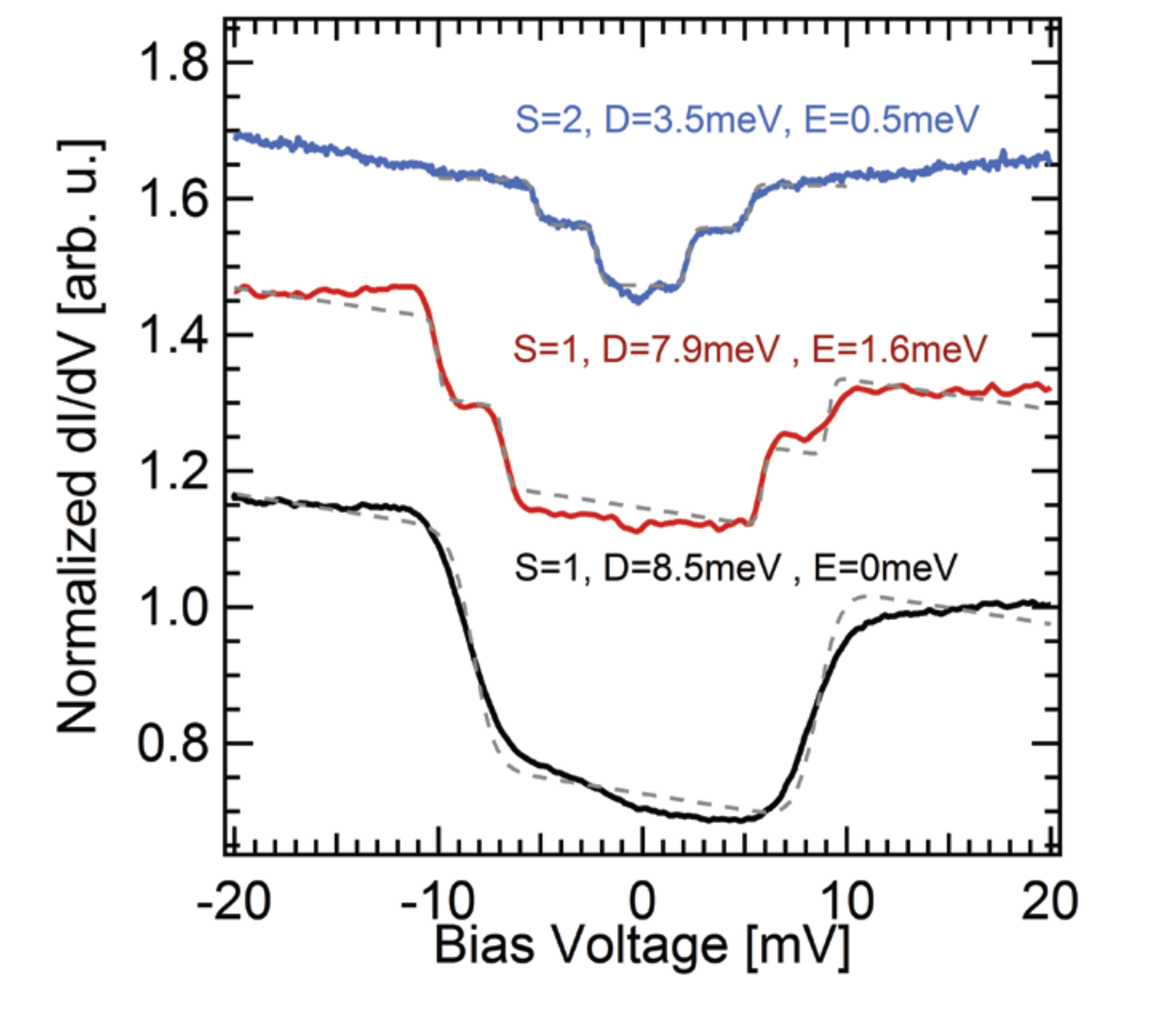}
\caption{ Comparison of the magnetic properties of all FeTPyP structures. dI/dV spectra recorded in the center of the FeTPyP molecules exhibit different numbers of steps symmetric in energy around the Fermi level. The spin
state corresponding to the fit (gray dashed lines) is given next to the graphs, as well as the resulting anisotropy
parameters $D$ and $E$. Feedback opened at: \SI{50}{\mV}, \SI{2}{\nA} with $V_{\text{mod}}=\SI{0.5}{\mV}$, T$ = \SI{4.8}{\kelvin}$ (black); \SI{50}{\mV},
\SI{1}{\nA} with $V_{\text{mod}}=\SI{0.25}{\mV}$,  T$ = \SI{1.1}{\kelvin}$ and B$= \SI{0.5}{\tesla}$ (to improve stability of measurements; see SI) (red and blue). }
\label{Fig2}
\end{figure}

%%%%%%%%%%%%%%%%%%%%%%%%%%%%%%%%%%%%%%%%%%%%%%%%%%%%%%%%%%%%%%%%%%%%%%%%%%%%%%%%%%%%%%%%%%%%%%%%%%%%%%%%%%%%%%%%%%%%%%%%%%%%%%%%%%%%%%%%%%%%%%%%%%%%%%%        Spectra       %%%%%%%%%%%%%%%%%%%%%%%%%%%%%%%%%%%%%%%%%%%%%%%%%%%%%%%%%%%%%%%%%%%%%%%%%%%%%%%%%%%%%%%%%%%%%%%%%%%%%%%%%%%%%%%%%%%%%%%%%%%%%%%%%%%%%%%%%%%%%%%%%%%%%%%%%%%%%%%%%%
To investigate the magnetic properties we recorded dI/dV spectra in the center of all types of molecules (Fig.\,\ref{Fig2}). In the densely-packed structure (black), all the molecules exhibit the same spectral features, consisting of one pair of steps, which is symmetric in energy around the Fermi level. 
 In the staggered structure, two types of molecules can be distinguished by their dI/dV spectra, as indicated by the blue and red colors. Both of these types exhibit two pairs of steps around the Fermi level. However, the steps differ considerably in their respective energies. 
All of these steps indicate inelastic tunneling processes and are attributed to spin excitations \cite{heinrich_single_atom_2004,tsukahara_adsorption_induced_2009,heinrich_change_2013,ohta_enhancement_2013,carmen_rubio_verdu_orbital_selective_2017,liu_large_2017,li_survival_2018}. 
Note that the different intensities at opposite bias polarities stem from a finite potential scattering term, which favors the electron-like excitations \cite{carmen_rubio_verdu_orbital_selective_2017}.

Spin excitations in the absence of an external magnetic field can only occur if the degeneracy of states with different spin projection quantum number are lifted by an anisotropic environment. 
The simplest model accounting for magnetic anisotropy is the spin Hamiltonian 
 ${\mathcal{H}_{\text{s}}=D\hat{S}^{2}_{\text{z}}+E(\hat{S}^{2}_{\text{x}}-\hat{S}^{2}_{\text{y}})
}$, with $D$ and $E$ denoting the axial and transverse anisotropy parameters and $\hat{S}^{2}_{\text{x,y,z}}$ the spin components in $x$,$y$ and $z$ direction, respectively. 
To obtain a qualitative picture of the magnetic anisotropy in the different molecular structures, we employ this spin Hamiltonian and its fit as implemented in Ref. \cite{markus_ternes_spin_2015}. 

For the spectra of the densely-packed molecules (black), we assume an intermediate-spin state of $S=1$, which is typical for Fe$^{2+}$ in square-planar ligand fields and has been observed for other Fe-porphyrins \cite{Ben_FeOEP_2015, bernien_fe-porphyrin_2007, wende_substrate-induced_2007, ohta_enhancement_2013,carmen_rubio_verdu_orbital_selective_2017}.
The spin excitation reflects an axial anisotropy energy $D$, which lifts the degeneracy between the $M_{\text{s}}=0$ and $M_{\text{s}}=\pm1$ spin projections. Without transverse anisotropy ($E=0$), this results in the emergence of one pair of steps at an energy corresponding to the axial anisotropy parameter $D=\SI{8.5}{\meV}$. 

The spectra of molecules in the staggered arrangement are qualitatively different. Both types of molecules exhibit two pairs of inelastic steps. 
The first type of spectra (red) can be reproduced within a $S=1$ ground state, including an axial anisotropy $D$ of similar size as for those of the molecules in the densely-packed arrangement, but with an additional transverse anisotropy parameter $E\neq0$. The latter reflects a new set of non-degenerate eigenstates arising from the symmetric and anti-symmetric linear combination of the $M_{\text{s}}=\pm1$ states. Hence, two inelastic excitations can occur, which are visible at energies $D\pm E/2$.

For the second type of molecules within the staggered arrangement (blue spectrum), the assumption of an intermediate S=1 spin state does not lead to reasonable fit results, as $D$ and $E$ would exhibit a  comparable magnitude, which does not fulfill the criterion of $E< D/3$ \cite{gatteschi_molecular_2006}. Instead, a high-spin state of $S=2$ including transverse anisotropy reproduces best the shape of the spectra. 

%%%%%%%%%%%%%%%%%%%%%%%%%%%%%%%%%%%%%%%%%%%%%%%%%%%%%%%%%%%%%%%%%%%%%%%%%%%%%%%%%%%  Delta F maps      %%%%%%%%%%%%%%%%%%%%%%%%%%%%%%%%%%%%%%%%%%%%%%%%%%%%%%%%%%%%%%%%%%%%%%%%%%%%%%%%%%%%%%%%%%%%%%%%%%

\begin{figure*}[ht]
\includegraphics[width=0.95\linewidth]{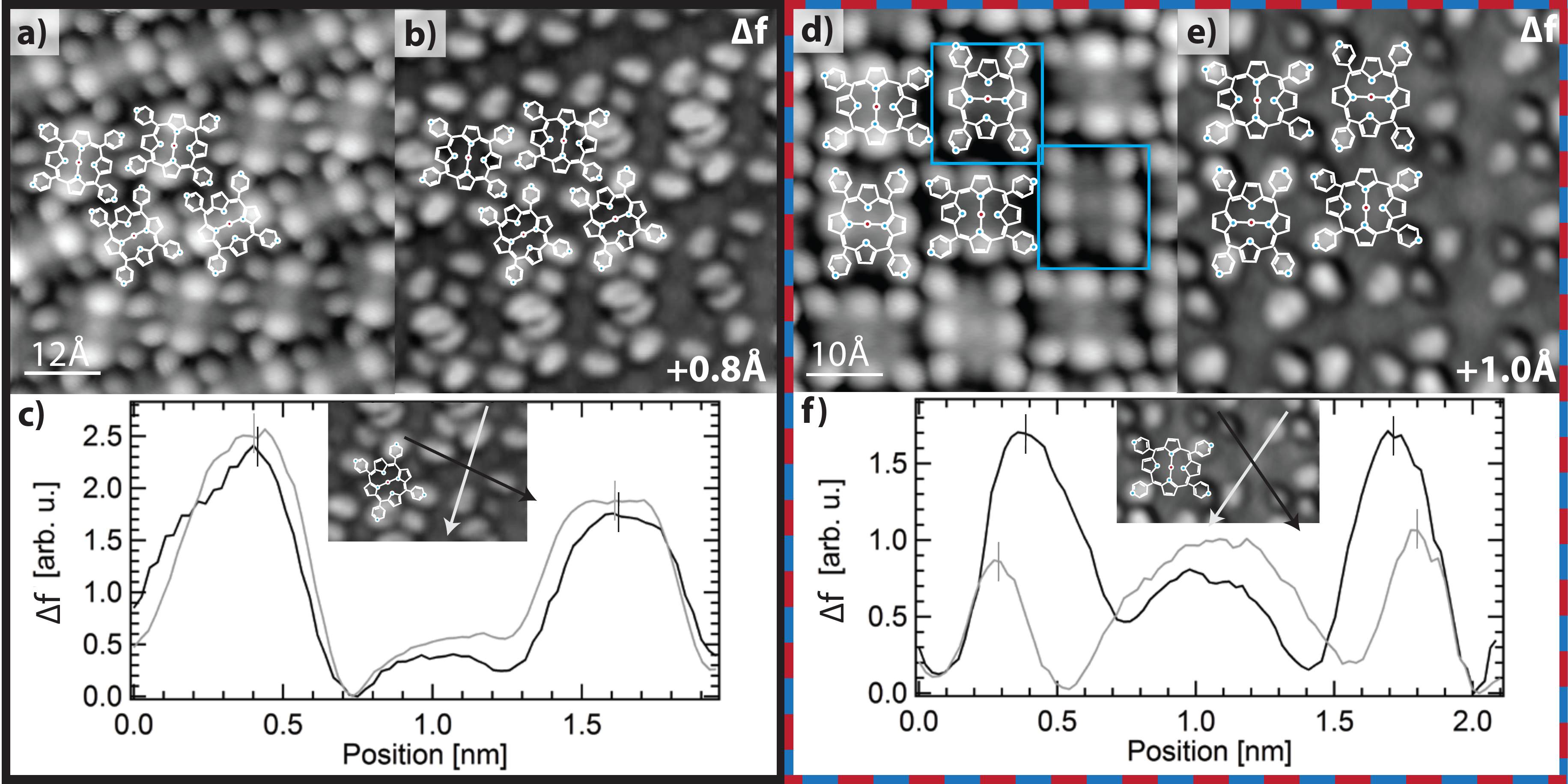}
\caption{Topographies and constant-height $\Delta$f maps recorded with a Cl-functionalized tip for a, b) the densely-packed structure and d, e) the distorted staggered structure. Whereas in the densely-packed structure, all pyridyl legs exhibit a similar rotational angle, in the staggered arrangement, two opposing pyridyl groups are rotated more strongly than the other two. The blue boxes mark the S=2 molecules. c, f) Line profile across the two diagonal axes showing the $\Delta$f signal on the pyridyl legs.  Topographies recorded at \SI{150}{\pico\A}, \SI{230}{\mV} (a) and \SI{160}{\pico\A}, \SI{230}{\mV} (d), respectively. For the recording of the $\Delta$f maps, the tip was approached by \SI{0.8}{\angstrom} (b) and \SI{1.0}{\angstrom} (e), as indicated in the respective images. c) All pyridyl legs exhibit the same rotational angle. The height difference between the pyridyl legs on top and bottom part of the molecule is  due to an overall inclination of the molecule following the corrugation of the herringbone reconstruction of the underlying Au(111) surface. f) In the staggered structure, the pyridyl legs along one diagonal are higher than along the other diagonal. The distances between the pyridyl rings along the molecular diagonals are also different, reflecting a lateral distortion of the molecule. }
\label{Fig3}
\end{figure*}

To understand the origin of the differences in the magnetic properties of the FeTPyP molecules, we performed non-contact atomic force microscopy measurements, using a qPlus tuning-fork-based sensor. Figure \ref{Fig3}a shows the STM topography of an island of molecules in the densely-packed arrangement together with the corresponding $\Delta$f map depicted in Fig.\,\ref{Fig3}b, which was recorded with a Cl-functionalized tip  \cite{Gross1110}. By comparing with the overlaid structure model, the pyridyl groups are identified as the most protruding moieties of the FeTPyP molecule. Moreover, at this tip-sample distance, only a faint trace of the contrast originating from the up-bending of two pyrrole groups is visible (see supplementary information (SI) for maps at different heights). The line profiles of the diagonal cross sections of one FeTPyP molecule show similar height of all pyridyl legs (Fig.\,\ref{Fig3}c). The small height difference of the pyridyl legs originates from the corrugation of the herringbone reconstruction of the underlying Au(111) substrate.

\begin{figure}[h]
\includegraphics[width=0.95\columnwidth]{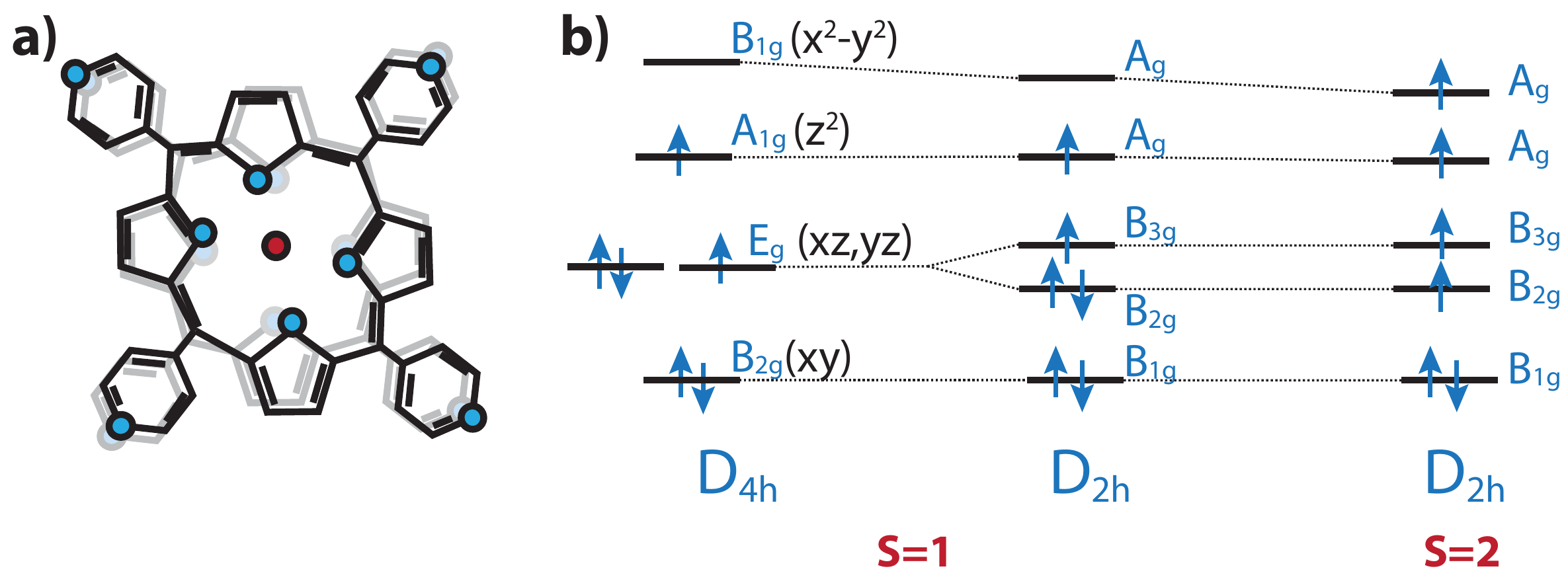}
\caption{a) Sketch of the molecular deformation in the staggered arrangement. The lateral deformation of the macrocycle induces a change of the Fe--N bond lengths, which reduces the symmetry of the ligand field. b) Proposed d-level configuration for the densely-packed molecules in a ligand field of $D_{4h}$ symmetry (left) and the molecules in a staggered arrangement with $D_{2h}$ symmetry of the ligand field (middle, right). The reduction of the ligand field symmetry induces a splitting of the d$_{xz}$ and d$_{yz}$ orbitals, which introduces a transverse magnetic anisotropy to the system. Upon increase of the Fe--N bond lengths, the d$_{x^2-y^2}$ orbital shifts down, eventually leading to a spin crossover.}
\label{Fig4}
\end{figure}

Next, we investigate the molecules in the staggered structure (Fig.\,\ref{Fig3}d and e). The S=1 and S=2 molecules cannot be distinguished in the $\Delta$f map (the S=2 molecules are marked by blue boxes). For both types of molecules in this arrangement, the pyridyl groups are the moiety that protrudes the most from the substrate, similar to the structure of the densely-packed arrangement. However, when analyzing the heights of the different pyridyl groups of one molecule, a symmetry breaking becomes apparent: the two pyridyl groups along one diagonal appear higher than the other two pyridyl groups (Fig. \ref{Fig3}f). The different heights reflect different rotational angles of the pairs. Moreover, the distance between the pyridyl groups is different along the two diagonal directions, as indicated by the vertical lines in the line profiles. The distance between the lower pyridyl groups (grey curve) is larger than the distance between the higher pyridyl groups (black curve). For different rotational angles between the pyridyl groups, simple geometry arguments (see SI) suggest an apparent length difference, as upon rotation of the pyridyl groups, the most protruding H atoms of the pyridyl groups are rotated away from their shortest connection line. For the hypothetical case with  one pair of pyridyl legs being perpendicular to the molecular plane, the center of the H atoms is above the center of the pyridyl ring, and the minimal distance of \SI{12.7}{\angstrom} is expected. For the other extreme case of the pair of pyridyl rings being almost flat, the H atoms of the oppositely rotated pyridyl legs are further apart by \SI{0.7}{\angstrom} (see Fig.~S2a). This sets an upper bound for the length difference one can obtain simply due to different rotational angles. In contrast, for the nine molecules in Fig.\,\ref{Fig3}e the experimentally determined length difference amounts to an average value of \SI{2.5}{\angstrom} (see Fig.~S2b). This deviation can only be explained by a lateral asymmetric distortion of the molecules. We suggest that the distortion of the molecules in this structure originates from intermolecular interactions of the molecules, where neighboring pyridyl groups arrange in an edge-to-face fashion (see SI for further details). This motif enhances the interaction between neighboring pyridyl groups, which leads to a stabilization of the molecular arrangement.
Importantly, this distortion is different from the saddle deformation of the molecule, which affects the overall out-of-plane molecular shape, but presumably preserves the positions of the N atoms with respect to the Fe center. As a consequence, the ligand field still exhibits $D_{4h}$ symmetry in the saddle configuration \cite{liu_large_2017}. The asymmetric distortion that we determine from the analysis of the line profiles of the staggered molecules now occurs in addition to the saddle-shape deformation. This leads to a lateral displacement of the N atoms of the pyrrole groups, which in turn reduces the molecular ligand field of the Fe center from $D_{4h}$ to $D_{2h}$ symmetry (see Fig.\,\ref{Fig4}a). This latter reduction of the symmetry seems to be responsible for the transverse magnetic anisotropy that we exclusively find in the distorted molecules in the staggered arrangement. Furthermore, we note that the reduction of the ligand field symmetry can be understood without any influence of the substrate. This is in contrast to previous studies that reported transverse anisotropies, where the atomic-scale adsorption site of the Fe on the substrate induced a symmetry breaking \cite{tsukahara_adsorption_induced_2009,Ben_FeOEP_2015}.

Having correlated the experimental observations of molecular distortions with the occurrence of magnetic anisotropy, we now provide a simple model in terms of the corresponding d-levels.
Multiple possibilities for the d-level occupations have been reported for FeTPP and FeTPyP molecules, with almost the same energy and resulting total spin $S=1$  \cite{carmen_rubio_verdu_orbital_selective_2017, vancoillie_multiconfigurational_2011, liao_electronic_2007,tanaka_electron_1986}. 
One possible d-level alignment for the ligand field of $D_{4h}$ symmetry, as it is assumed for the molecules in the densely-packed structure, and the corresponding level occupation are shown in Fig.\,\ref{Fig4}b.
Such a d-level occupation has also been obtained from DFT calculations of FeTPP on Au(111) \cite{carmen_rubio_verdu_orbital_selective_2017}. Due to the similarity of FeTPP to FeTPyP in the densely-packed structure (black) in terms of both the molecular saddle shape and the dI/dV spectra, we expect that their findings also apply to our system.
In this d-level scheme, the d$_{x^2-y^2}$ orbital remains unoccupied, as it lies highest in energy due to its direct overlap with the states of the neighboring N atoms in the adjacent pyrrole groups. The d$_{z^2}$ orbital also has a component inside this $xy$ plane defined by the four N atoms and is therefore only singly occupied. The d$_{xz}$ and d$_{yz}$ orbitals are degenerate and share three electrons. The doubly occupied d$_{xy}$ orbital is least affected by the ligand field as it has all its components between the N atoms. This occupation leads to a spin state of S=1 and, upon introducing spin-orbit coupling, to finite values of the axial anisotropy energy $D$. 
In the staggered arrangement, in contrast, the reduction of the ligand field symmetry towards $D_{2h}$ lifts the degeneracy between the d$_{xz}$ and d$_{yz}$ orbitals, which induces a transverse anisotropy in the system. This explains the emergence of the second pair of steps in the dI/dV spectra of Fig.\,\ref{Fig2}. Hence, the different magnetic anisotropy parameters observed for the molecules in the different arrangements can be understood from the symmetry of their ligand fields as directly seen by the lateral distortion of the molecule.

As mentioned above, no difference between the S=1 and S=2 molecules in the staggered structure is detectable in terms of the molecular distortion in any of our STM and AFM measurements. 
In general, a spin crossover towards a high-spin state becomes energetically favorable for elongated Fe--N bonds \cite{liu_large_2017,wang_intramolecularly_2015} and has been observed for other FeTPP and FeTPyP complexes on Au(111) \cite{liu_large_2017,kuang_mechanically_controlled_2017}. Due to the smaller influence of the ligand field on the central Fe atom, larger Fe--N bond lengths lead to energetically closer d levels that can hence all be populated by electrons (see Fig.\,\ref{Fig4}b). 
DFT calculations indicate that changes of the Fe--N bond length in the order of several \si{\pm} are sufficient to alter the ground state of Fe porphyrins \cite{bhandary_graphene_2011,bhandary_manipulation_2013,bhandary_correlated_2016}. However, changes of such a length scale are beyond the resolution of our measurements. Another possible explanation is the elongation of the Fe--N bonds in the direction towards the surface due to stronger interactions of the Fe atom with the substrate. This, however, is also not detectable with the STM or AFM. 

In conclusion, we have shown that different adsorption structures of FeTPyP exist on the Au(111) substrate, which lead to different molecular configurations and magnetic states. We could identify different lateral distortions of the molecular macrocycle. We relate these distortions to the fingerprints in the magnetic excitation spectra. When the symmetry of the ligand field is maintained, the molecules lie in a $S=1$ state with axial anisotropy, but negligible transverse anisotropy. In the other structure, an asymmetric lateral distortion of the macrocycle reduces the symmetry of the ligand field to $D_{2h}$, introducing transverse anisotropy as a response to the lifting of the d$_{xz}$ and d$_{yz}$ orbital degeneracies. Our results directly visualize the interplay of molecular distortions and magnetic properties. Furthermore, they reflect the possibility of tuning magnetic anisotropy energies by concise molecular distortions.

Funding by the ERC Consolidator grant "NanoSpin" as well as the International Max Planck Research School on "Functional Interfaces in Physics and Chemistry" is gratefully acknowledged.

\bibliographystyle{apsrev4-1}
%\bibliography{paper}

\begin{thebibliography}{99}
\providecommand{\latin}[1]{#1}
\makeatletter
\providecommand{\doi}
  {\begingroup\let\do\@makeother\dospecials
  \catcode`\{=1 \catcode`\}=2 \doi@aux}
\providecommand{\doi@aux}[1]{\endgroup\texttt{#1}}
\makeatother
\providecommand*\mcitethebibliography{\thebibliography}
\csname @ifundefined\endcsname{endmcitethebibliography}
  {\let\endmcitethebibliography\endthebibliography}{}
%\begin{mcitethebibliography}{34}
\providecommand*\natexlab[1]{#1}
\providecommand*\mciteSetBstSublistMode[1]{}
\providecommand*\mciteSetBstMaxWidthForm[2]{}
\providecommand*\mciteBstWouldAddEndPuncttrue
  {\def\EndOfBibitem{\unskip.}}
\providecommand*\mciteBstWouldAddEndPunctfalse
  {\let\EndOfBibitem\relax}
\providecommand*\mciteSetBstMidEndSepPunct[3]{}
\providecommand*\mciteSetBstSublistLabelBeginEnd[3]{}
\providecommand*\EndOfBibitem{}
\mciteSetBstSublistMode{f}
\mciteSetBstMaxWidthForm{subitem}{(\alph{mcitesubitemcount})}
\mciteSetBstSublistLabelBeginEnd
  {\mcitemaxwidthsubitemform\space}
  {\relax}
  {\relax}

\bibitem[Rau \latin{et~al.}(2014)Rau, Baumann, Rusponi, Donati, Stepanow,
  Gragnaniello, Dreiser, Piamonteze, Nolting, Gangopadhyay, Albertini,
  Macfarlane, Lutz, Jones, Gambardella, Heinrich, and Brune]{rau_reaching_2014}
Rau,~I.~G.; Baumann,~S.; Rusponi,~S.; Donati,~F.; Stepanow,~S.;
  Gragnaniello,~L.; Dreiser,~J.; Piamonteze,~C.; Nolting,~F.; Gangopadhyay,~S.;
  Albertini,~O.~R.; Macfarlane,~R.~M.; Lutz,~C.~P.; Jones,~B.~A.;
  Gambardella,~P.; Heinrich,~A.~J.; Brune,~H. Reaching the magnetic anisotropy
  limit of a 3d metal atom. \emph{Science} \textbf{2014}, \emph{344},
  988--992\relax
\mciteBstWouldAddEndPuncttrue
\mciteSetBstMidEndSepPunct{\mcitedefaultmidpunct}
{\mcitedefaultendpunct}{\mcitedefaultseppunct}\relax
\EndOfBibitem
\bibitem[Heinrich \latin{et~al.}(2015)Heinrich, Braun, Pascual, and
  Franke]{Ben_FeOEP_2015}
Heinrich,~B.~W.; Braun,~L.; Pascual,~J.~I.; Franke,~K.~J. Tuning the magnetic
  anisotropy of single molecules. \emph{Nano Lett.} \textbf{2015}, \emph{15},
  4024--4028\relax
\mciteBstWouldAddEndPuncttrue
\mciteSetBstMidEndSepPunct{\mcitedefaultmidpunct}
{\mcitedefaultendpunct}{\mcitedefaultseppunct}\relax
\EndOfBibitem
\bibitem[Hiraoka \latin{et~al.}(2017)Hiraoka, Takagi, Arafune, Tsukahara,
  Watanabe, Kawai, and Takagi]{hiraoka_single_molecule_2017}
Hiraoka,~R.; Takagi,~N.; Arafune,~R.; Tsukahara,~N.; Watanabe,~S.; Kawai,~M.;
  Takagi,~N. Single-molecule quantum dot as a {Kondo} simulator. \emph{Nat.
  Commun.} \textbf{2017}, \emph{8}, 16012\relax
\mciteBstWouldAddEndPuncttrue
\mciteSetBstMidEndSepPunct{\mcitedefaultmidpunct}
{\mcitedefaultendpunct}{\mcitedefaultseppunct}\relax
\EndOfBibitem
\bibitem[Ormaza \latin{et~al.}(2017)Ormaza, Abufager, Verlhac, Bachellier,
  Bocquet, Lorente, and Limot]{ormaza_controlled_2017}
Ormaza,~M.; Abufager,~P.; Verlhac,~B.; Bachellier,~N.; Bocquet,~M.-L.;
  Lorente,~N.; Limot,~L. Controlled spin switching in a metallocene molecular
  junction. \emph{Nat. Commun.} \textbf{2017}, \emph{8}, 1974\relax
\mciteBstWouldAddEndPuncttrue
\mciteSetBstMidEndSepPunct{\mcitedefaultmidpunct}
{\mcitedefaultendpunct}{\mcitedefaultseppunct}\relax
\EndOfBibitem
\bibitem[Chen \latin{et~al.}(2018)Chen, Isshiki, Baretzky, Balashov, and
  Wulfhekel]{chen_abrupt_2018}
Chen,~J.; Isshiki,~H.; Baretzky,~C.; Balashov,~T.; Wulfhekel,~W. Abrupt
  {Switching} of {Crystal} {Fields} during {Formation} of {Molecular}
  {Contacts}. \emph{ACS Nano} \textbf{2018}, \emph{12}, 3280--3286\relax
\mciteBstWouldAddEndPuncttrue
\mciteSetBstMidEndSepPunct{\mcitedefaultmidpunct}
{\mcitedefaultendpunct}{\mcitedefaultseppunct}\relax
\EndOfBibitem
\bibitem[Wang \latin{et~al.}(2015)Wang, Pang, Kuang, Shi, Shang, Liu, and
  Lin]{wang_intramolecularly_2015}
Wang,~W.; Pang,~R.; Kuang,~G.; Shi,~X.; Shang,~X.; Liu,~P.~N.; Lin,~N.
  Intramolecularly resolved {Kondo} {Resonance} of {High-Spin}
  {Fe(II)}-porphyrin adsorbed on {Au}(111). \emph{Phys. Rev. B} \textbf{2015},
  \emph{91}, 045440\relax
\mciteBstWouldAddEndPuncttrue
\mciteSetBstMidEndSepPunct{\mcitedefaultmidpunct}
{\mcitedefaultendpunct}{\mcitedefaultseppunct}\relax
\EndOfBibitem
\bibitem[Rubio-Verd\'{u} \latin{et~al.}(2018)Rubio-Verd\'{u}, Sarasola, Choi,
  Majzik, Ebeling, Calvo, Ugeda, Garcia-Lekue, S\'{a}nchez-Portal, and
  Pascual]{carmen_rubio_verdu_orbital_selective_2017}
Rubio-Verd\'{u},~C.; Sarasola,~A.; Choi,~D.-J.; Majzik,~Z.; Ebeling,~R.;
  Calvo,~M.~R.; Ugeda,~M.~M.; Garcia-Lekue,~A.; S\'{a}nchez-Portal,~D.;
  Pascual,~J.~I. Orbital-selective spin excitation of a magnetic porphyrin.
  \emph{Comms. Phys.} \textbf{2018}, \emph{1}, 15\relax
\mciteBstWouldAddEndPuncttrue
\mciteSetBstMidEndSepPunct{\mcitedefaultmidpunct}
{\mcitedefaultendpunct}{\mcitedefaultseppunct}\relax
\EndOfBibitem
\bibitem[Liu \latin{et~al.}(2017)Liu, Fu, Guan, Shao, Meng, Guo, and
  Wang]{liu_large_2017}
Liu,~B.; Fu,~H.; Guan,~J.; Shao,~B.; Meng,~S.; Guo,~J.; Wang,~W. An
  {Iron}-{Porphyrin} {Complex} with {Large} {Easy}-{Axis} {Magnetic}
  {Anisotropy} on {Metal} {Substrate}. \emph{ACS Nano} \textbf{2017},
  \emph{11}, 11402--11408\relax
\mciteBstWouldAddEndPuncttrue
\mciteSetBstMidEndSepPunct{\mcitedefaultmidpunct}
{\mcitedefaultendpunct}{\mcitedefaultseppunct}\relax
\EndOfBibitem
\bibitem[Auw\"arter \latin{et~al.}(2007)Auw\"arter, Weber-Bargioni, Brink,
  Riemann, Schiffrin, Ruben, and Barth]{auwarter_controlled_2007}
Auw\"arter,~W.; Weber-Bargioni,~A.; Brink,~S.; Riemann,~A.; Schiffrin,~A.;
  Ruben,~M.; Barth,~J.~V. Controlled Metalation of Self-Assembled Porphyrin
  Nanoarrays in Two Dimensions. \emph{Chem. Phys. Chem.} \textbf{2007},
  \emph{8}, 250--254\relax
\mciteBstWouldAddEndPuncttrue
\mciteSetBstMidEndSepPunct{\mcitedefaultmidpunct}
{\mcitedefaultendpunct}{\mcitedefaultseppunct}\relax
\EndOfBibitem
\bibitem[Chen \latin{et~al.}(2017)Chen, Shulai, Lotze, Czekelius, Paulus, and
  Franke]{xianwen_chen_conformational_2017}
Chen,~X.; Shulai,~L.; Lotze,~C.; Czekelius,~C.; Paulus,~B.; Franke,~K.~J.
  Conformational adaptation and manipulation of manganese
  tetra(4-pyridyl)porphyrin molecules on Cu(111). \emph{J. Chem. Phys.}
  \textbf{2017}, \emph{146}, 092316\relax
\mciteBstWouldAddEndPuncttrue
\mciteSetBstMidEndSepPunct{\mcitedefaultmidpunct}
{\mcitedefaultendpunct}{\mcitedefaultseppunct}\relax
\EndOfBibitem
\bibitem[Fleischer \latin{et~al.}(1971)Fleischer, Palmer, Srivastava, and
  Chatterjee]{Fetpyp}
Fleischer,~E.~B.; Palmer,~J.~M.; Srivastava,~T.~S.; Chatterjee,~A. Thermodyamic
  and kinetic properties of an iron-porphyrin system. \emph{J. Am. Chem. Soc.}
  \textbf{1971}, \emph{93}, 3162--3167\relax
\mciteBstWouldAddEndPuncttrue
\mciteSetBstMidEndSepPunct{\mcitedefaultmidpunct}
{\mcitedefaultendpunct}{\mcitedefaultseppunct}\relax
\EndOfBibitem
\bibitem[Nabid \latin{et~al.}(2009)Nabid, Zamiraei, Sedghi, and
  Safari]{Fetpyp2}
Nabid,~M.~R.; Zamiraei,~Z.; Sedghi,~R.; Safari,~N. Cationic metalloporphyrins
  for synthesis of conducting, water-soluble polyaniline. \emph{React. Funct.
  Polym.} \textbf{2009}, \emph{69}, 319--324\relax
\mciteBstWouldAddEndPuncttrue
\mciteSetBstMidEndSepPunct{\mcitedefaultmidpunct}
{\mcitedefaultendpunct}{\mcitedefaultseppunct}\relax
\EndOfBibitem
\bibitem[Heinrich \latin{et~al.}(2013)Heinrich, Ahmadi, Müller, Braun,
  Pascual, and Franke]{heinrich_change_2013}
Heinrich,~B.~W.; Ahmadi,~G.; Müller,~V.~L.; Braun,~L.; Pascual,~J.~I.;
  Franke,~K.~J. Change of the Magnetic Coupling of a Metal–Organic Complex
  with the Substrate by a Stepwise Ligand Reaction. \emph{Nano Lett.}
  \textbf{2013}, \emph{13}, 4840--4843\relax
\mciteBstWouldAddEndPuncttrue
\mciteSetBstMidEndSepPunct{\mcitedefaultmidpunct}
{\mcitedefaultendpunct}{\mcitedefaultseppunct}\relax
\EndOfBibitem
\bibitem[Gopakumar \latin{et~al.}(2012)Gopakumar, Tang, Morillo, and
  Berndt]{gopakumar_transfer_2012}
Gopakumar,~T.~G.; Tang,~H.; Morillo,~J.; Berndt,~R. Transfer of {Cl} {Ligands}
  between {Adsorbed} {Iron} {Tetraphenylporphyrin} {Molecules}. \emph{J. Am.
  Chem. Soc.} \textbf{2012}, \emph{134}, 11844--11847\relax
\mciteBstWouldAddEndPuncttrue
\mciteSetBstMidEndSepPunct{\mcitedefaultmidpunct}
{\mcitedefaultendpunct}{\mcitedefaultseppunct}\relax
\EndOfBibitem
\bibitem[Albrecht \latin{et~al.}(2016)Albrecht, Bischoff, Auw\"arter, Barth,
  and Repp]{albrecht_direct_2016}
Albrecht,~F.; Bischoff,~F.; Auw\"arter,~W.; Barth,~J.~V.; Repp,~J. Direct
  Identification and Determination of Conformational Response in Adsorbed
  Individual Nonplanar Molecular Species Using Noncontact Atomic Force
  Microscopy. \emph{Nano Lett.} \textbf{2016}, \emph{16}, 7703--7709\relax
\mciteBstWouldAddEndPuncttrue
\mciteSetBstMidEndSepPunct{\mcitedefaultmidpunct}
{\mcitedefaultendpunct}{\mcitedefaultseppunct}\relax
\EndOfBibitem
\bibitem[Auw\"arter \latin{et~al.}(2007)Auw\"arter, Klappenberger,
  Weber-Bargioni, Schiffrin, Strunskus, Wöll, Pennec, Riemann, and
  Barth]{auwarter_conformational_2007}
Auw\"arter,~W.; Klappenberger,~F.; Weber-Bargioni,~A.; Schiffrin,~A.;
  Strunskus,~T.; Wöll,~C.; Pennec,~Y.; Riemann,~A.; Barth,~J.~V.
  Conformational Adaptation and Selective Adatom Capturing of
  Tetrapyridyl-porphyrin Molecules on a Copper (111) Surface. \emph{J. Am.
  Chem. Soc.} \textbf{2007}, \emph{129}, 11279--11285\relax
\mciteBstWouldAddEndPuncttrue
\mciteSetBstMidEndSepPunct{\mcitedefaultmidpunct}
{\mcitedefaultendpunct}{\mcitedefaultseppunct}\relax
\EndOfBibitem
\bibitem[Zotti \latin{et~al.}(2007)Zotti, Teobaldi, Hofer, Auw\"arter,
  Weber-Bargioni, and Barth]{zotti_ab-initio_2007}
Zotti,~L.; Teobaldi,~G.; Hofer,~W.; Auw\"arter,~W.; Weber-Bargioni,~A.;
  Barth,~J. Ab-initio calculations and {STM} observations on tetrapyridyl and
  Fe({II})-tetrapyridyl-porphyrin molecules on Ag(111). \emph{Surf. Sci.}
  \textbf{2007}, \emph{601}, 2409--2414\relax
\mciteBstWouldAddEndPuncttrue
\mciteSetBstMidEndSepPunct{\mcitedefaultmidpunct}
{\mcitedefaultendpunct}{\mcitedefaultseppunct}\relax
\EndOfBibitem
\bibitem[Heinrich \latin{et~al.}(2004)Heinrich, Gupta, Lutz, and
  Eigler]{heinrich_single_atom_2004}
Heinrich,~A.~J.; Gupta,~J.~A.; Lutz,~C.~P.; Eigler,~D.~M. Single-{Atom}
  {Spin}-{Flip} {Spectroscopy}. \emph{Science} \textbf{2004}, \emph{306},
  466--469\relax
\mciteBstWouldAddEndPuncttrue
\mciteSetBstMidEndSepPunct{\mcitedefaultmidpunct}
{\mcitedefaultendpunct}{\mcitedefaultseppunct}\relax
\EndOfBibitem
\bibitem[Tsukahara \latin{et~al.}(2009)Tsukahara, Noto, Ohara, Shiraki, Takagi,
  Takata, Miyawaki, Taguchi, Chainani, Shin, and
  Kawai]{tsukahara_adsorption_induced_2009}
Tsukahara,~N.; Noto,~K.-i.; Ohara,~M.; Shiraki,~S.; Takagi,~N.; Takata,~Y.;
  Miyawaki,~J.; Taguchi,~M.; Chainani,~A.; Shin,~S.; Kawai,~M.
  Adsorption-{Induced} {Switching} of {Magnetic} {Anisotropy} in a {Single}
  {Iron}({II}) {Phthalocyanine} {Molecule} on an {Oxidized} {Cu}(110)
  {Surface}. \emph{Phys. Rev. Lett.} \textbf{2009}, \emph{102}, 167203\relax
\mciteBstWouldAddEndPuncttrue
\mciteSetBstMidEndSepPunct{\mcitedefaultmidpunct}
{\mcitedefaultendpunct}{\mcitedefaultseppunct}\relax
\EndOfBibitem
\bibitem[Ohta \latin{et~al.}(2013)Ohta, Arafune, Tsukahara, Kawai, and
  Takagi]{ohta_enhancement_2013}
Ohta,~N.; Arafune,~R.; Tsukahara,~N.; Kawai,~M.; Takagi,~N. Enhancement of
  Inelastic Electron Tunneling Conductance Caused by Electronic Decoupling in
  Iron Phthalocyanine Bilayer on Ag(111). \emph{J. Phys. Chem. C}
  \textbf{2013}, \emph{117}, 21832--21837\relax
\mciteBstWouldAddEndPuncttrue
\mciteSetBstMidEndSepPunct{\mcitedefaultmidpunct}
{\mcitedefaultendpunct}{\mcitedefaultseppunct}\relax
\EndOfBibitem
\bibitem[Li \latin{et~al.}(2018)Li, Merino-Díez, Carbonell-Sanromà,
  Vilas-Varela, de~Oteyza, Peña, Corso, and Pascual]{li_survival_2018}
Li,~J.; Merino-Díez,~N.; Carbonell-Sanromà,~E.; Vilas-Varela,~M.;
  de~Oteyza,~D.~G.; Peña,~D.; Corso,~M.; Pascual,~J.~I. Survival of Spin State
  in Magnetic Porphyrins Contacted by Graphene Nanoribbons. \emph{Sci. Adv.}
  \textbf{2018}, \emph{4}, eaaq0582\relax
\mciteBstWouldAddEndPuncttrue
\mciteSetBstMidEndSepPunct{\mcitedefaultmidpunct}
{\mcitedefaultendpunct}{\mcitedefaultseppunct}\relax
\EndOfBibitem
\bibitem[{Markus Ternes}(2015)]{markus_ternes_spin_2015}
{Markus Ternes}, Spin excitations and correlations in scanning tunneling
  spectroscopy. \emph{New J. Phys.} \textbf{2015}, \emph{17}, 063016\relax
\mciteBstWouldAddEndPuncttrue
\mciteSetBstMidEndSepPunct{\mcitedefaultmidpunct}
{\mcitedefaultendpunct}{\mcitedefaultseppunct}\relax
\EndOfBibitem
\bibitem[Bernien \latin{et~al.}(2007)Bernien, Xu, Miguel, Piantek, Eckhold,
  Luo, Kurde, Kuch, Baberschke, Wende, and
  Srivastava]{bernien_fe-porphyrin_2007}
Bernien,~M.; Xu,~X.; Miguel,~J.; Piantek,~M.; Eckhold,~P.; Luo,~J.; Kurde,~J.;
  Kuch,~W.; Baberschke,~K.; Wende,~H.; Srivastava,~P. Fe-porphyrin monolayers
  on ferromagnetic substrates: Electronic structure and magnetic coupling
  strength. \emph{Phys. Rev. B} \textbf{2007}, \emph{76}, 214406\relax
\mciteBstWouldAddEndPuncttrue
\mciteSetBstMidEndSepPunct{\mcitedefaultmidpunct}
{\mcitedefaultendpunct}{\mcitedefaultseppunct}\relax
\EndOfBibitem
\bibitem[Wende \latin{et~al.}(2007)Wende, Bernien, Luo, Sorg, Ponpandian,
  Kurde, Miguel, Piantek, Xu, Eckhold, Kuch, Baberschke, Panchmatia, Sanyal,
  Oppeneer, and Eriksson]{wende_substrate-induced_2007}
Wende,~H.; Bernien,~M.; Luo,~J.; Sorg,~C.; Ponpandian,~N.; Kurde,~J.;
  Miguel,~J.; Piantek,~M.; Xu,~X.; Eckhold,~P.; Kuch,~W.; Baberschke,~K.;
  Panchmatia,~P.~M.; Sanyal,~B.; Oppeneer,~P.~M.; Eriksson,~O.
  Substrate-induced magnetic ordering and switching of iron porphyrin
  molecules. \emph{Nat. Mater.} \textbf{2007}, \emph{6}, 516--520\relax
\mciteBstWouldAddEndPuncttrue
\mciteSetBstMidEndSepPunct{\mcitedefaultmidpunct}
{\mcitedefaultendpunct}{\mcitedefaultseppunct}\relax
\EndOfBibitem
\bibitem[Gatteschi \latin{et~al.}(2006)Gatteschi, Sessoli, and
  Villain]{gatteschi_molecular_2006}
Gatteschi,~D.; Sessoli,~R.; Villain,~J. \emph{Molecular nanomagnets}; Oxford
  University Press, 2006\relax
\mciteBstWouldAddEndPuncttrue
\mciteSetBstMidEndSepPunct{\mcitedefaultmidpunct}
{\mcitedefaultendpunct}{\mcitedefaultseppunct}\relax
\EndOfBibitem
\bibitem[Gross \latin{et~al.}(2009)Gross, Mohn, Moll, Liljeroth, and
  Meyer]{Gross1110}
Gross,~L.; Mohn,~F.; Moll,~N.; Liljeroth,~P.; Meyer,~G. The Chemical Structure
  of a Molecule Resolved by Atomic Force Microscopy. \emph{Science}
  \textbf{2009}, \emph{325}, 1110--1114\relax
\mciteBstWouldAddEndPuncttrue
\mciteSetBstMidEndSepPunct{\mcitedefaultmidpunct}
{\mcitedefaultendpunct}{\mcitedefaultseppunct}\relax
\EndOfBibitem
\bibitem[Vancoillie \latin{et~al.}(2011)Vancoillie, Zhao, Tran, Hendrickx, and
  Pierloot]{vancoillie_multiconfigurational_2011}
Vancoillie,~S.; Zhao,~H.; Tran,~V.~T.; Hendrickx,~M. F.~A.; Pierloot,~K.
  Multiconfigurational Second-Order Perturbation Theory Restricted Active Space
  ({RASPT}2) Studies on Mononuclear First-Row Transition-Metal Systems.
  \emph{J. Chem. Theory Comput.} \textbf{2011}, \emph{7}, 3961--3977\relax
\mciteBstWouldAddEndPuncttrue
\mciteSetBstMidEndSepPunct{\mcitedefaultmidpunct}
{\mcitedefaultendpunct}{\mcitedefaultseppunct}\relax
\EndOfBibitem
\bibitem[Liao \latin{et~al.}(2007)Liao, Watts, and Huang]{liao_electronic_2007}
Liao,~M.-S.; Watts,~J.~D.; Huang,~M.-J. Electronic Structure of Some
  Substituted Iron({II}) Porphyrins. Are They Intermediate or High Spin?
  \emph{J. Phys. Chem. A} \textbf{2007}, \emph{111}, 5927--5935\relax
\mciteBstWouldAddEndPuncttrue
\mciteSetBstMidEndSepPunct{\mcitedefaultmidpunct}
{\mcitedefaultendpunct}{\mcitedefaultseppunct}\relax
\EndOfBibitem
\bibitem[Tanaka \latin{et~al.}(1986)Tanaka, Elkaim, Li, Jue, Coppens, and
  Landrum]{tanaka_electron_1986}
Tanaka,~K.; Elkaim,~E.; Li,~L.; Jue,~Z.~N.; Coppens,~P.; Landrum,~J. Electron
  density studies of porphyrins and phthalocyanines. {IV}. Electron density
  distribution in crystals of (meso-tetraphenylporphinato) iron({II}). \emph{J.
  Chem. Phys.} \textbf{1986}, \emph{84}, 6969--6978\relax
\mciteBstWouldAddEndPuncttrue
\mciteSetBstMidEndSepPunct{\mcitedefaultmidpunct}
{\mcitedefaultendpunct}{\mcitedefaultseppunct}\relax
\EndOfBibitem
\bibitem[Kuang \latin{et~al.}(2017)Kuang, Zhang, Lin, Pang, Shi, Xu, and
  Lin]{kuang_mechanically_controlled_2017}
Kuang,~G.; Zhang,~Q.; Lin,~T.; Pang,~R.; Shi,~X.; Xu,~H.; Lin,~N.
  Mechanically-{Controlled} {Reversible} {Spin} {Crossover} of {Single}
  {Fe}-{Porphyrin} {Molecules}. \emph{ACS Nano} \textbf{2017}, \emph{11},
  6295--6300\relax
\mciteBstWouldAddEndPuncttrue
\mciteSetBstMidEndSepPunct{\mcitedefaultmidpunct}
{\mcitedefaultendpunct}{\mcitedefaultseppunct}\relax
\EndOfBibitem
\bibitem[Bhandary \latin{et~al.}(2011)Bhandary, Ghosh, Herper, Wende, Eriksson,
  and Sanyal]{bhandary_graphene_2011}
Bhandary,~S.; Ghosh,~S.; Herper,~H.; Wende,~H.; Eriksson,~O.; Sanyal,~B.
  Graphene as a {Reversible} {Spin} {Manipulator} of {Molecular} {Magnets}.
  \emph{Phy. Rev. Lett.} \textbf{2011}, \emph{107}, 257202\relax
\mciteBstWouldAddEndPuncttrue
\mciteSetBstMidEndSepPunct{\mcitedefaultmidpunct}
{\mcitedefaultendpunct}{\mcitedefaultseppunct}\relax
\EndOfBibitem
\bibitem[Bhandary \latin{et~al.}(2013)Bhandary, Brena, Panchmatia, Brumboiu,
  Bernien, Weis, Krumme, Etz, Kuch, Wende, Eriksson, and
  Sanyal]{bhandary_manipulation_2013}
Bhandary,~S.; Brena,~B.; Panchmatia,~P.~M.; Brumboiu,~I.; Bernien,~M.;
  Weis,~C.; Krumme,~B.; Etz,~C.; Kuch,~W.; Wende,~H.; Eriksson,~O.; Sanyal,~B.
  Manipulation of spin state of iron porphyrin by chemisorption on magnetic
  substrates. \emph{Phys. Rev. B} \textbf{2013}, \emph{88}, 024401\relax
\mciteBstWouldAddEndPuncttrue
\mciteSetBstMidEndSepPunct{\mcitedefaultmidpunct}
{\mcitedefaultendpunct}{\mcitedefaultseppunct}\relax
\EndOfBibitem
\bibitem[Bhandary \latin{et~al.}(2016)Bhandary, Sch\"uler, Thunstr\"om,
  di~Marco, Brena, Eriksson, Wehling, and Sanyal]{bhandary_correlated_2016}
Bhandary,~S.; Sch\"uler,~M.; Thunstr\"om,~P.; di~Marco,~I.; Brena,~B.;
  Eriksson,~O.; Wehling,~T.; Sanyal,~B. Correlated electron behavior of
  metal-organic molecules: {Insights} from density functional theory combined
  with many-body effects using exact diagonalization. \emph{Phys. Rev. B}
  \textbf{2016}, \emph{93}, 155158\relax
\mciteBstWouldAddEndPuncttrue
\mciteSetBstMidEndSepPunct{\mcitedefaultmidpunct}
{\mcitedefaultendpunct}{\mcitedefaultseppunct}\relax
\EndOfBibitem
%
\end{thebibliography}

\begin{thebibliography}{10}
\providecommand{\latin}[1]{#1}
\makeatletter
\providecommand{\doi}
  {\begingroup\let\do\@makeother\dospecials
  \catcode`\{=1 \catcode`\}=2 \doi@aux}
\providecommand{\doi@aux}[1]{\endgroup\texttt{#1}}
\makeatother
\providecommand*\mcitethebibliography{\thebibliography}
\csname @ifundefined\endcsname{endmcitethebibliography}
  {\let\endmcitethebibliography\endthebibliography}{}
%\begin{mcitethebibliography}{5}
\providecommand*\natexlab[1]{#1}
\providecommand*\mciteSetBstSublistMode[1]{}
\providecommand*\mciteSetBstMaxWidthForm[2]{}
\providecommand*\mciteBstWouldAddEndPuncttrue
  {\def\EndOfBibitem{\unskip.}}
\providecommand*\mciteBstWouldAddEndPunctfalse
  {\let\EndOfBibitem\relax}
\providecommand*\mciteSetBstMidEndSepPunct[3]{}
\providecommand*\mciteSetBstSublistLabelBeginEnd[3]{}
\providecommand*\EndOfBibitem{}
\mciteSetBstSublistMode{f}
\mciteSetBstMaxWidthForm{subitem}{(\alph{mcitesubitemcount})}
\mciteSetBstSublistLabelBeginEnd
  {\mcitemaxwidthsubitemform\space}
  {\relax}
  {\relax}

\bibitem[sof(accessed June, 2018)]{Ssoftware}
{Probe Particle Model, http://nanosurf.fzu.cz/ppr/}. accessed June, 2018\relax
\mciteBstWouldAddEndPuncttrue
\mciteSetBstMidEndSepPunct{\mcitedefaultmidpunct}
{\mcitedefaultendpunct}{\mcitedefaultseppunct}\relax
\EndOfBibitem
\bibitem[Hapala \latin{et~al.}(2014)Hapala, Kichin, Wagner, Tautz, Temirov, and
  Jel\'inek]{Shapala_mechanism_2014}
Hapala,~P.; Kichin,~G.; Wagner,~C.; Tautz,~F.~S.; Temirov,~R.; Jel\'inek,~P.
  Mechanism of high-resolution {STM}/{AFM} imaging with functionalized tips.
  \emph{Phys. Rev. B} \textbf{2014}, \emph{90}, 085421\relax
\mciteBstWouldAddEndPuncttrue
\mciteSetBstMidEndSepPunct{\mcitedefaultmidpunct}
{\mcitedefaultendpunct}{\mcitedefaultseppunct}\relax
\EndOfBibitem
\bibitem[Hapala \latin{et~al.}(2014)Hapala, Temirov, Tautz, and
  Jel\'inek]{Shapala_origin_2014}
Hapala,~P.; Temirov,~R.; Tautz,~F.~S.; Jel\'inek,~P. Origin of
  {High}-{Resolution} {IETS}-{STM} {Images} of {Organic} {Molecules} with
  {Functionalized} {Tips}. \emph{Phys. Rev. Lett.} \textbf{2014}, \emph{113},
  226101\relax
\mciteBstWouldAddEndPuncttrue
\mciteSetBstMidEndSepPunct{\mcitedefaultmidpunct}
{\mcitedefaultendpunct}{\mcitedefaultseppunct}\relax
\EndOfBibitem
\bibitem[Gross \latin{et~al.}(2012)Gross, Mohn, Moll, Schuler, Criado, Guitian,
  Pena, Gourdon, and Meyer]{Sgross_bond_order_2012}
Gross,~L.; Mohn,~F.; Moll,~N.; Schuler,~B.; Criado,~A.; Guitian,~E.; Pena,~D.;
  Gourdon,~A.; Meyer,~G. Bond-{Order} {Discrimination} by {Atomic} {Force}
  {Microscopy}. \emph{Science} \textbf{2012}, \emph{337}, 1326--1329\relax
\mciteBstWouldAddEndPuncttrue
\mciteSetBstMidEndSepPunct{\mcitedefaultmidpunct}
{\mcitedefaultendpunct}{\mcitedefaultseppunct}\relax
\EndOfBibitem
	\end{thebibliography}

\clearpage

\setcounter{figure}{0}
\setcounter{section}{0}
\setcounter{equation}{0}
\renewcommand{\theequation}{S\arabic{equation}}
\renewcommand{\thefigure}{S\arabic{figure}}

\onecolumngrid

\renewcommand{\Fig}[1]{\mbox{Fig.\unitspace\ref{Sfig:#1}}}
\renewcommand{\Figure}[1]{\mbox{Figure\unitspace\ref{Sfig:#1}}}

\newcommand{\vsigma}{\mbox{\boldmath $\sigma$}}

\section*{\Large{Supplementary Information}}

\vspace{0.7cm}
\section{Constant-height $\Delta$f maps at various tip-sample distances}
\begin{figure*}[ht]
\includegraphics[width=0.95\linewidth]{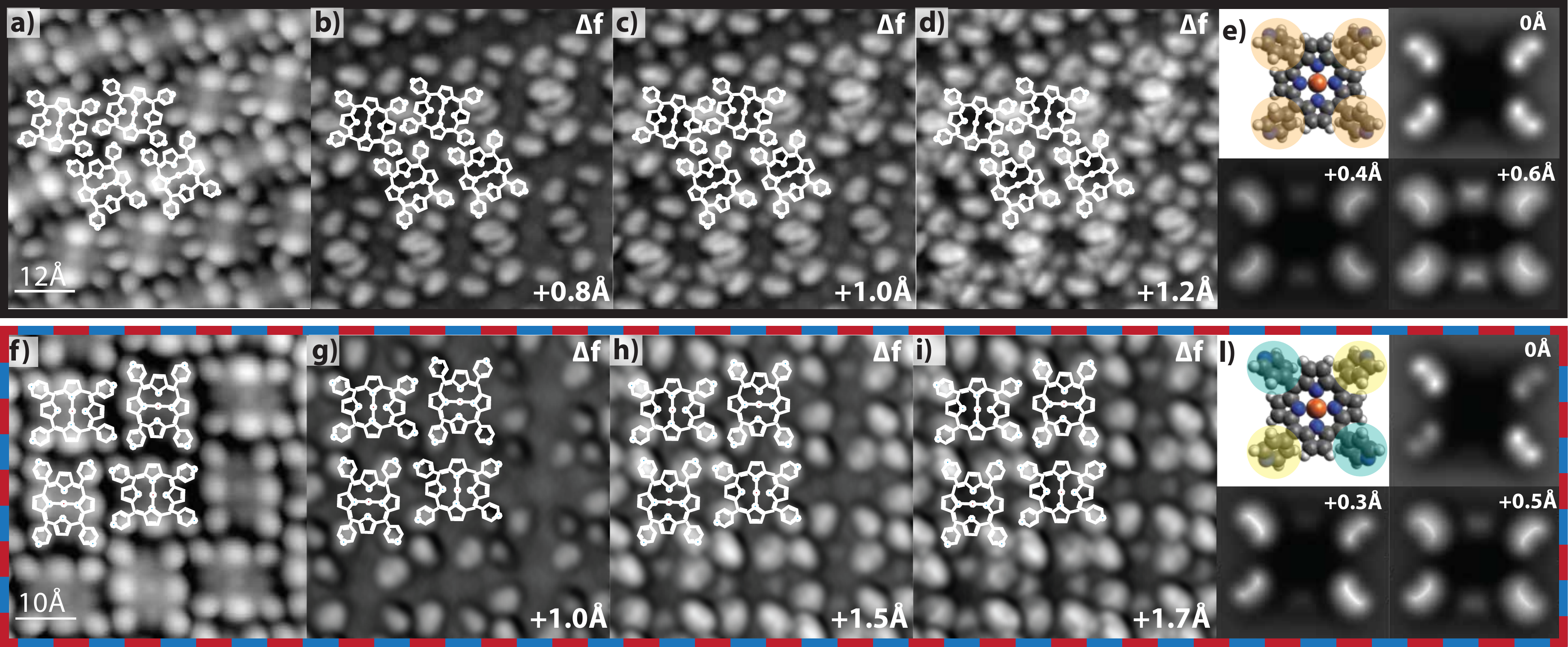}
\caption{STM topographies and constant-height $\Delta$f maps recorded with a functionalized Cl tip at different tip-sample distances. a-d) of the densely-packed structure; f-i) of the staggered arrangement. At larger tip-sample distances, only the pyridyl groups are visible. Upon further tip approach, a contrast arises from the upper pyrrole groups of the organic macrocyle.  Topographies recorded at \SI{150}{\pico\A}, \SI{230}{\mV} (a) and \SI{160}{\pico\A}, \SI{230}{\mV} (d), respectively. For the $\Delta$f maps, the tip position with respect to this setpoint is indicated in the respective images. d,h) Sketch of the molecular structures and simulated $\Delta$f maps for different tip-sample distances.}
\label{SupplFig1}
\end{figure*}

The Lennard-Jones like dependence of the frequency shift on the tip-sample distance does not allow to unambiguously determine the heights of molecular moieties from a single $\Delta$f image. Therefore, $\Delta$f maps recorded at several tip-sample distances are shown in Fig.\,\ref{SupplFig1} for both the densely-packed structure (b-d) and the staggered arrangement (g-i). The change in contrast confirms the interpretation of the pyridyl legs to be the highest molecular moiety in both structures. At larger tip-sample distances, only the pyridyl legs can be identified (\ref{SupplFig1}b,g).
Upon further tip approach, the two upper pyrrole groups of the saddle-shape deformation of the molecule can also be  distinguished (\ref{SupplFig1}c,h).

To conceive an idea of the size of the molecular deformations with respect to the rotational angles of the pyridyl groups in both arrangements, Fig.\,\ref{SupplFig1}e,l show sketches of the molecular structures in gas phase, together with simulated images of the resulting $\Delta$f maps.
For both structures, the dihedral angles that define the rotation of the pyridyl legs with respect to the surface plane were manually varied and fixed during the DFT minimization. The best agreement between the resulting simulations of the $\Delta$f maps based on the DFT calculations and experimental findings was obtained for a dihedral angle of \SI{45}{\degree} for the densely-packed structure and angles of \SI{44}{\degree} and \SI{49}{\degree} for the distorted staggered molecules. The simulations of the $\Delta$f maps were performed with a Cl tip for different tip-sample distances by using the simulation program of Hapala and coworkers \cite{Ssoftware}, which is based on the force field between a flexible tip and the molecules \cite{Shapala_mechanism_2014,Shapala_origin_2014}.
 In agreement with the height-dependent $\Delta$f maps at larger tip separations, only the pyridyl legs are distinguishable. For the chosen molecular structures, the $\Delta$f maps agree with the experimentally observed symmetries, showing similar contrast on all four pyridyl legs for the densely-packed structure and a symmetry breaking for the molecules of the staggered arrangement. Upon closer tip approach, also the upper pyrrole groups of the macrocycle become apparent. Despite of the simulations being performed for a manually distorted molecule in gas phase, they give an indication of the small size of intramolecular distortions that suffice to influence the magnetic properties.

\section{Analysis of the molecular distortions}

\begin{figure*}[ht]
\includegraphics[width=0.95\linewidth]{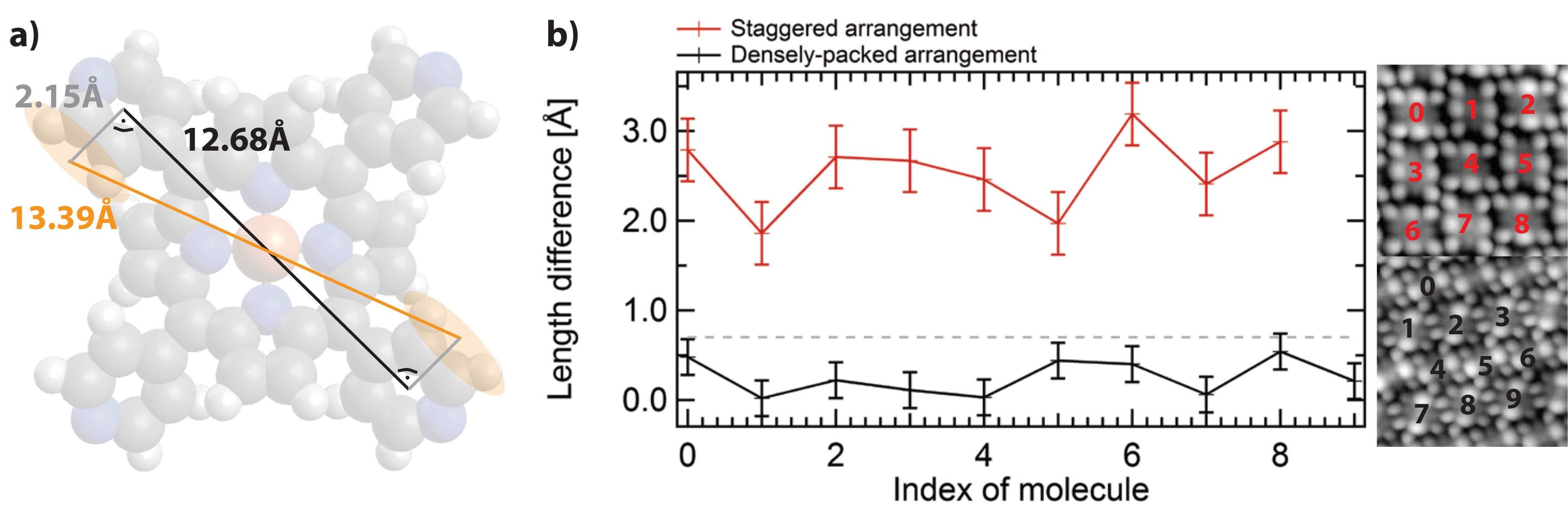}
\caption{a) Origin of the apparent length difference for pyridyl groups with different rotational angles. For completely perpendicular pyridyl groups (black line), the length between the pyridyl groups is minimized. For any flatter arrangement of the pyridyl groups, the rotation of the H atoms of the pyridyl groups away from the diagonal will increase the apparent length (orange line). b) Experimentally determined difference between the length of the opposing pyridyl groups along the different diagonals for both molecular structures. Whereas hardly any difference can be observed for the molecules in the densely-packed arrangement, a clear asymmetry is apparent for the staggered arrangement. The grey dashed line corresponds to the upper bound of the length difference of \SI{0.7}{\angstrom} in the absence of intramolecular distortions. The topographies on the right show the corresponding molecules that were considered in the determination of the distortions.}
\label{SupplFig3}
\end{figure*}

In the $\Delta$f maps in Fig.~3 and Fig.\,\ref{SupplFig1}, bright lobes are apparent at the positions of the pyridyl groups. From the analysis of the line profiles of the molecules in the staggered arrangement, an average length difference of \SI{2.5}{\angstrom} was determined between the pyridyl groups of different rotational angles (Fig.\,\ref{SupplFig3}b). As the pyridyl moieties are rotated by a certain angle out of the molecular plane, the H atoms of the pyridyl ring are the most protruding atoms and responsible for the contrast. For the hypothetical case of perpendicular pyridyl groups, the distance between two opposing pyridyl groups is minimized and amounts to \SI{12.68}{\angstrom} (black line in Fig.\,\ref{SupplFig3}a), as can be determined from DFT calculations. If the pyridyl groups adapt a flatter rotational angle, the opposing rotation direction of the pyridyl legs leads to an increase of the measured distance (see orange line), whose maximal length $l$ can be calculated for the hypothetical case of (almost) flat pyridyl legs as:
\begin{equation}
l=\sqrt{(\SI{12.68}{\angstrom})^2+(\SI{2.15}{\angstrom})^2}=\SI{13.39}{\angstrom}.
\end{equation}
Therefore, the apparent length difference arising from asymmetrically rotated pyridyl legs can at most amount to \SI{0.7}{\angstrom}. Any larger difference, as observed for the molecules in the staggered arrangement, hence, originates from intramolecular distortions.

Note that the use of flexible tips for AFM imaging has led to the observation of apparent bond elongations in molecular complexes \cite{Sgross_bond_order_2012,Shapala_mechanism_2014}. In our case, however, a tip relaxation should have a similar effect along both diagonals of the molecule and should not be restricted to only one diagonal. Hence, we exclude imaging artefacts as the cause of the observed length difference.

\section{Origin of the molecular distortions}
\begin{figure*}[ht]
\includegraphics[width=0.7\linewidth]{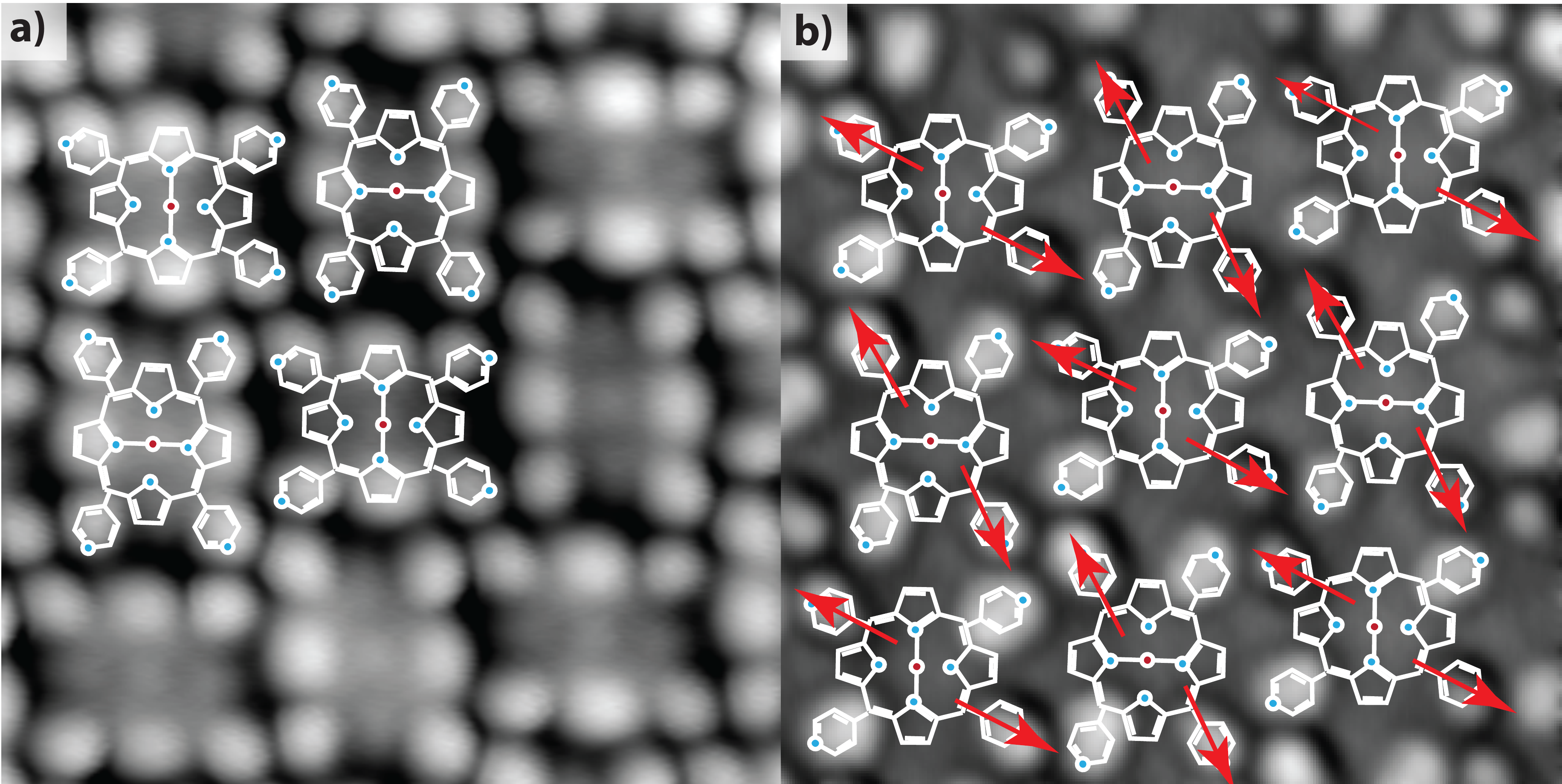}
\caption{a) STM topography and b) constant-height $\Delta$f map, showing the arrangement of the FeTPyP molecules. The red arrows indicate the lower pyridyl groups, which coordinate to the pyridyl groups of the neighboring molecule, thus introducing a small diagonal intramolecular distortion. The topography was recorded at \SI{160}{\pico\A}, \SI{230}{\mV}; for the $\Delta$f map, the tip was approached by \SI{1}{\angstrom} with respect to this setpoint.  }
\label{SupplFig4}
\end{figure*}

The self-assembly of the FeTPyP molecules on Au(111) is driven by the interplay of intermolecular as well as molecule-substrate interactions. The occurrence of different structures reflects that the precise balance of these interactions determines the resulting structure. As shown in the main manuscript, AFM measurements reveal a distortion of the molecule in the staggered arrangement, whereas the molecules appear undistorted in the densely-packed arrangement. 

 In Fig.\,\ref{SupplFig4}, the molecular arrangement in the  staggered structure is shown, with neighboring molecules being rotated by 90$^\circ$ with respect to the direction of their saddle shape. The red arrows  in Fig.\,\ref{SupplFig4}b indicate the flatter pair of pyridyl rings, for which an elongation has been oberved. As observable from the overlaid structure model, the flatter pyridyl groups point towards the upstanding pyridyl group of the neighboring molecule in an egde-to-face fashion. By introducing a small diagonal distortion within the molecules, the distance between the neighboring pyridyl groups is reduced, such that the intermolecular structure is stabilized.

\section{B-field dependence of the spin excitations}

\begin{figure*}[ht]
\includegraphics[width=0.95\linewidth]{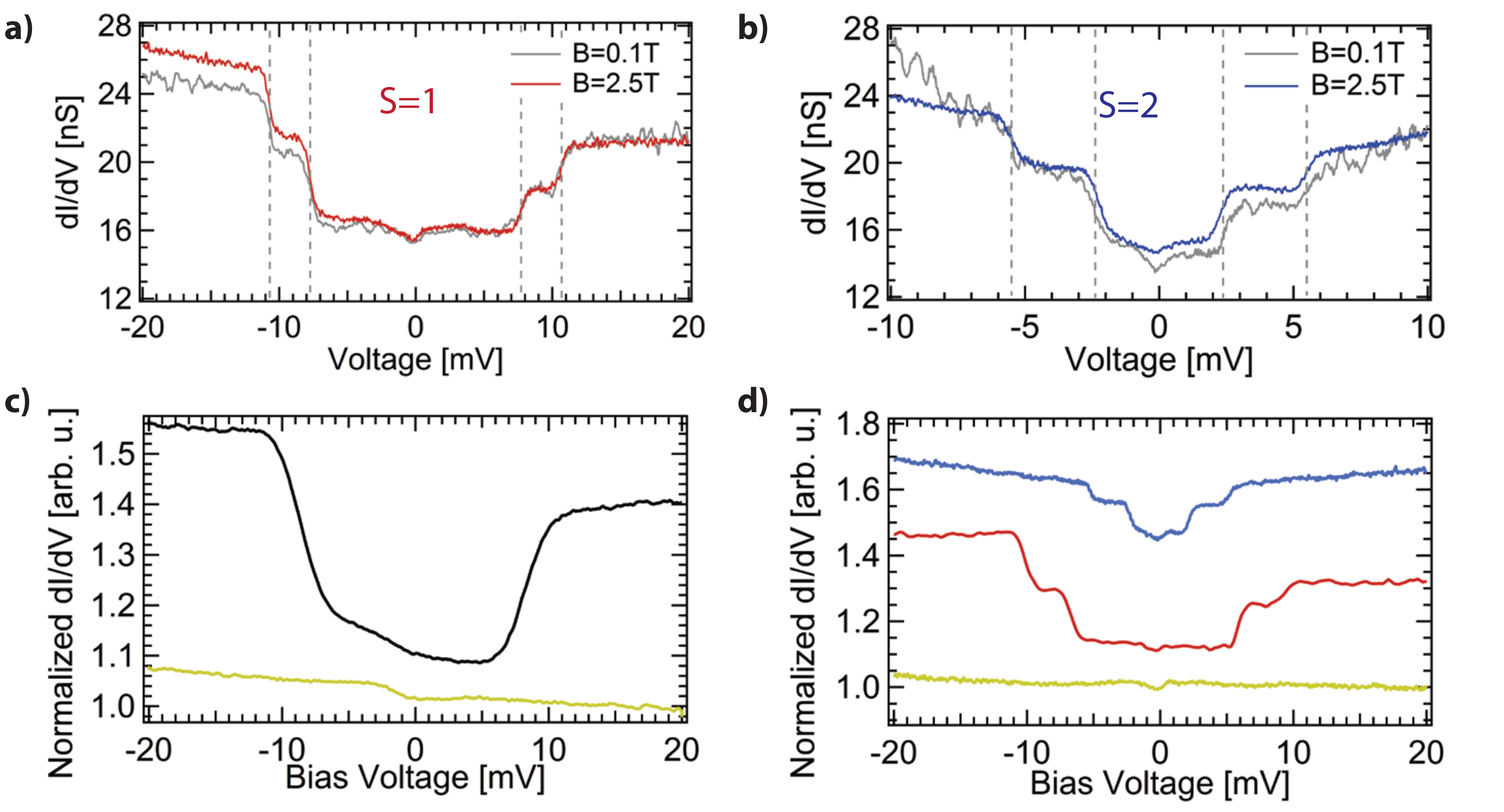}
\caption{
a,b) dI/dV spectra recorded on the two types of molecules of the staggered structure at two different magnetic fields of B$= \SI{0.1}{\tesla}$ and B$= \SI{2.5}{\tesla}$. Feedback opened at \SI{50}{\mV},
\SI{1}{\nA} with $V_{\text{mod}}=\SI{0.25}{\mV}$, T$ = \SI{1.1}{\kelvin}$.
c,d) dI/dV spectra of Fig.~2 together with their respective reference spectrum recorded on the Au(111) substrate, indicating that the dip-like feature at the Fermi level is a tip artefact and does not have a magnetic origin. c) Feedback opened at  \SI{50}{\mV}, \SI{2}{\nA} with $V_{\text{mod}}=\SI{0.5}{\mV}$, T$ = \SI{4.8}{\kelvin}$. d) Feedback opened at \SI{50}{\mV},
\SI{1}{\nA} with $V_{\text{mod}}=\SI{0.25}{\mV}$, B$= \SI{0.5}{\tesla}$, T$ = \SI{1.1}{\kelvin}$.}
\label{SupplFig2}
\end{figure*}

To confirm the magnetic origin of the steps in the differential conductance, dI/dV spectra were recorded
at different magnetic fields perpendicular to the surface. In Fig.\,\ref{SupplFig2}a,b, dI/dV spectra recorded at
\SI{0.1}{\tesla} and \SI{2.5}{\tesla} are depicted for the S=1 and S=2 molecules, respectively. For neither of the two
molecules, a shift of the step position with increasing B field is visible. On the first glance, the absence of a shift of the step position with higher B fields seems to
contradict the interpretation of a magnetic origin of the steps. To estimate the expected shift of the
steps, we make use of the relation
\begin{equation}
\Delta E= g\mu_{\text{B}}BM_S
\end{equation} 
with $\mu_{\text{B}}$ denoting the Bohr magneton and g the gyromagnetic ratio, which is g=2 in case of a free
spin. The shift should therefore amount to approx. \SI{300}{\micro\V} at \SI{2.5}{\tesla}, which is in the order of the thermal
broadening at the measurement temperature of \SI{1.1}{\kelvin}. Moreover, in case of transverse anisotropy of
the system, the resulting intermixing of the states of opposite spin projection would counteract the
splitting of the levels with increasing B field such that the corresponding Zeeman shift would be even smaller.
Hence, the absence of a shift does not allow to interpret the origin of the steps as non-magnetic. 

As the quality of the spectra is better at finite B fields due to the enhanced mechanical stability of the STM and since no effect of a B field on
the molecular properties could be observed, the blue and red spectra shown in Fig.~2 were recorded at B = \SI{0.5}{\tesla}.

The spectra shown in Fig.~2 both on the densely-packed and the staggered arrangement all exhibit a small dip right at the Fermi level. To ensure that this feature is not of magnetic origin, the dI/dV spectra are shown again in Fig.\,\ref{SupplFig2}c,d together with the corresponding reference spectrum that was recorded on the Au(111) substrate. By comparison of the spectral shape around $E_F$, the dip can be assigned to a tip artefact and does not relate to a magnetic fingerprint.

\end{document}